\documentclass{article}
\usepackage{ijcai19}

\usepackage{times}
\usepackage{xcolor}
\usepackage{soul}
\usepackage[utf8]{inputenc}
\usepackage[small]{caption}
\usepackage{graphicx}
\usepackage{caption}
\usepackage{subcaption}
\usepackage{booktabs}
\usepackage{blindtext}
\usepackage{amsmath}
\usepackage{multirow}
\usepackage{color}
\usepackage{makecell}
\usepackage{longtable}
\usepackage{lipsum}

\title{Social Media-based User Embedding: A Literature Review}

\author{
Shimei Pan\footnote{Contact Author}
\and
Tao Ding\\
Department of Information Systems,University of Maryland, Baltimore County\\
\{shimei, taoding01\}@umbc.edu
}

\begin{document}

\maketitle

\begin{abstract}
Automated representation  learning is  behind many recent success stories in machine learning. It is often used to transfer knowledge learned from a large dataset (e.g., raw text) to tasks for which only a small number of training examples are available.  In this paper, we review recent advance in learning to represent social media users in low-dimensional embeddings.   The  technology  is  critical  for creating high performance social media-based human traits and behavior models since the ground truth for assessing latent human traits and behavior is often expensive to acquire at a large scale.  In this survey, we review typical methods for learning a unified user embeddings from heterogeneous user data  (e.g., combines social media texts with images to learn a unified user representation).  Finally we point out some current issues and future directions.

\end{abstract}
\section{Introduction}
People currently spend a significant amount of time on social media to express opinions, interact with friends and share ideas.  As a result, social media data contain rich information that is indicative of who we are and predictive of our online or real world behavior. With the recent advent of big data analytics, social media-based human trait and behavioral analytics has increasingly been used to better understand human minds and predict human behavior. Prior research has demonstrated that by analyzing the information in a user’s social media account, we can infer many latent user characteristics such as political leaning~\cite{pennacchiotti2011machine,kosinski2013private,benton2016learning}, brand preferences~\cite{pennacchiotti2011machine,yang2015using}, emotions~\cite{kosinski2013private}, mental disorders~\cite{de2013predicting}, personality~\cite{kosinski2013private,schwartz2013personality,liu2016analyzing,GolbeckRT11}, substance use~\cite{kosinski2013private,ding2017multi} and sexual orientation~\cite{kosinski2013private}. 

Although social media allow us to easily record a large amount of user data, the characteristics of social media data also bring significant challenges to automated data analytics. For example, the texts and images are unstructured data. Making sense of unstructured data is always a big challenge. It is also hard to efficiently search and analyze a large social graph. Moreover, social media analytics can easily suffer from the {\em curse of dimensionality problem}.  If we use the basic text features such as unigrams or TF*IDF scores as the features to represent text, we can easily have hundreds of thousands of text features. Moreover, assessing human traits and behavior often requires psychometric evaluations or medical diagnosis, which are expensive to perform at a large scale (e.g., only trained professionals can provide an accurate assessment on whether someone has substance use disorders or not).  Without proper user feature learning, a machine learning model can easily overfit the training data and will not generalize well to new data. 

Recent years have seen a surge in methods that automatically encode features in low-dimensional embeddings using techniques such as dimension reduction and deep learning~\cite{Mikolov2013,le2014,Bengio2013,grover2016node2vec,perozzi2014deepwalk}. {\em Representation learning} has increasingly become a critical tool to boost the performance of complex machine learning applications. In this paper, we review recent work on automatically learning user representations from social media data. Since automated user embedding simultaneously performs latent feature learning and dimension reduction, it can help downstream tasks to  avoid overfitting and boost performance. 



\section{Overview}
Here we define social media-based user embedding as the function that maps raw user features in a high dimensional space to dense vectors in a low dimensional embedding space. The learned user embeddings often  capture the essential characteristics of individuals on social media. Since they are quite general, the learned user embeddings can be used to support diverse downstream  user analysis tasks such as user preference prediction~\cite{pennacchiotti2011machine},  personality modeling~\cite{kosinski2013private}  and  depression  detection~\cite{amir2017quantifying}. 

Automated user embedding is different from traditional user feature extraction where a pre-defined set of features is extracted from data. For example, based on the Linguistic Inquiry and Word Count (LIWC) dictionary~\cite{pennebaker2015development},  a set of pcycholinguistic features can be extracted from text.   Similarly, a set of egocentric network features such as degree, size and betweenness centrality can be extracted from one's social network. The main difference between user embedding and traditional user feature extraction is that in user embedding, the user features are not pre-defined. They are latent features automatically learned from data.  

\begin{figure}[t]
\includegraphics[width=0.45\textwidth]{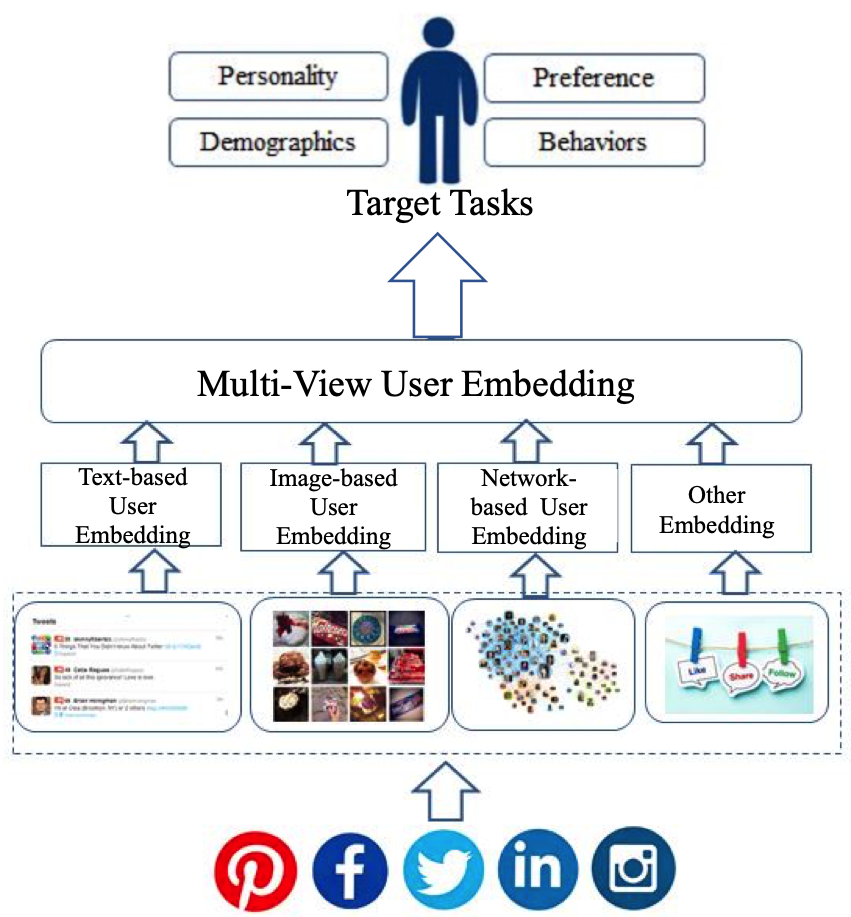}
\caption{A Typical System Architecture}
\label{fig:intro}
\end{figure}

Figure~\ref{fig:intro} shows the typical architecture of a system that employs automated user embedding for personal traits and behavior analysis. One or more types of user data are first extracted from a social media account.  For each type of user data such as text or image, a set of latent user features is learned  automatically via {\em single-view user embedding} (e.g., text-based user embedding and image-based user embedding).  The embeddings learned from different types of user data (e.g., text embeddings and image embeddings) are combined to form a single unified user representation via {\em multi-view user embedding}. The output of multi-view user embedding is then used in subsequent applications to predict human traits and behavior. 

Given the page limit, we define the scope of this survey quite narrowly to include only embedding methods published within the last 10 years that have been used to learn user representations from social media data. Although very relevant, We exclude embedding methods that do not learn a representation of social media users. For example, we exclude the papers on learning word embeddings from social media data~\cite{zeng2018}.    
Table~\ref{tab:MethodOverview} lists the papers included in our survey.  We summarize each paper along six dimensions: Data Type, Single-view Embedding Method, Auxiliary Training Task, Multi-view Embedding Method, Target Task and Supervised Tuning. 

Among them, {\em Data Type} is used  to indicate the types of social media data used in each study. Here, 
\textit{text} refers to user-generated text data (e.g., tweets or status update on Facebook);
\textit{like} refers to things/people a social media user likes such as books, celebrities, music, movies, TV shows, photos and products; 
\textit{user profile} includes demographic information (gender, age, occupation, relationship status etc.) and aggregated statistics   (the number of friends, followers, followings etc.); 
\textit{image} includes the profile and background photos as well as the images shared on social media; 
\textit{social network} refers to social connections between different user accounts such as the friendship network on Facebook and the follower/retweet network on Twitter. 

We also list the main methods used in {\em Single-view} and {\em Multi-view user embedding}.  They typically employ unsupervised or self-supervised learning to automatically uncover the latent structures/relations in the raw data. To employ self-supervised user embedding, frequently an {\em Auxiliary Training Task} is employed for which the system can easily construct a large number of training examples. We will explain the details of these methods later. {\em Target task}  describes the downstream applications that make use of the learned embeddings. We also indicate whether the learned embeddings are further tuned/adjusted so that they are optimized for the target tasks. 

In the following, We first present the typical Single-view User Embedding methods.   Then we summarize the methods that combine multiple types of user information together to form a unified user representation. 


\section{Single-View User Embedding}
Since most papers in table~\ref{tab:MethodOverview} learn user embedding from text, we focus on text-based user embeddings. We will also discuss the typical methods used in learning user embeddings from social networks. Finally, we briefly describe how to learn user embeddings from other types of social media data such as likes and images.

\begin{table*}[ht!]
\small
\centering
    \begin{tabular}{@{}p{2.4cm}p{1.5cm}p{2cm}p{2cm}p{2cm}p{3cm}p{1.5cm}@{}}
      \toprule
       \textbf{Paper} & \textbf{Data Type} &
       \textbf{Single-view embedding Method}& \textbf{Auxiliary Training Task} &\textbf{Multi-view embedding Task}
       &\textbf{Target Task} & \textbf{Target-task Tuning}\\
       \midrule
       \cite{pennacchiotti2011machine}&text,user profile & LDA & NA &concatenation & political learning&No \\
       &&&&& ethnicity\\
       &&&&& user preference\\\hline
       \cite{kosinski2013private}&like&SVD&NA&NA&age,gender, personality&No\\
       &&&&&relationship\\
       &&&&&substance use\\
       &&&&&religion\\
       &&&&&political learning\\\hline
       \cite{schwartz2013personality} &text & LDA & NA & NA & age,gender&No\\  
       &&&&&personality& \\\hline
       \cite{gao2014modeling}&text &SVD&NA&NA&attribute&No\\\hline 
       \cite{perozzi2014deepwalk} &network &DeepWalk & node prediction & NA & user interests & No\\\hline 
       \cite{preoctiuc2015analysis} &text, user profile&SVD  &NA&concatenation& occupation&No\\
       &&Word2Vec & word prediction & \\
       \hline 
       \cite{song2015interest} &text&LDA&NA&NA&interest&No\\\hline 
       \cite{hu2016language} &text&LDA &NA&NA&occupation&No\\\hline 
       \cite{song2016volunteerism} &text &LDA &NA&& volunteerism & Yes\\
       & network &&&& \\
       & user profile &&&& \\\hline 
       \cite{benton2016learning} &text&PCA&NA&CCA& topic engagement &No\\
       &&Word2vec& word prediction&&friend recommendation\\
       &network&PCA & NA & &age, gender\\
       &&&&&political learning\\\hline
       \cite{wallace2016modelling} &text&NA& word prediction & NA & sarcasm detection&Yes\\\hline 
       \cite{ding2017multi}&text&Doc2vec&word prediction &CCA&substance use& No\\
       &like&Doc2vec& like prediction && \\\hline
       \cite{preoctiuc2017beyond}&text &Word2vec&word prediction&NA&political learning&No\\
       && LDA & NA & 
       \\\hline 
      \cite{amir2017quantifying}&text &Word2vec & word prediction & NA &depression &No\\
      &&&& \\\hline 
     \cite{zhang2017user} &network&NA &NA&UPPSNE&gender&No\\
     &user profile&&&&education&\\\hline 
     \cite{wang2017community} & network & NMF&NA& NA& political learning &NA \\
        &&&&&community&\\\hline
                                                
     \cite{ding2018predicting} & like & SVD, LDA &NA & NA& delay discounting & No \\
                                &    & Doc2vec &  && \\\hline
      \cite{liao2018attributed} &network,user profile& NA & 
      NA & SNE & group classification &No \\\hline
     \cite{zhang2018user} &image &VGGNet & NA & NA & user image popularity & Yes\\
     &text & LSTM & NA & & \\\hline 
     \cite{ribeiro2018characterizing} &text &GloVe&NA&GraphSage& hateful user&No\\
     &network &  && & &\\\hline 
     \cite{zhang2018anrl} &network & NA & NA&ANRL & group classification &No \\
     &user profile&&&&& \\\hline
     \cite{do2018twitter} & text & Doc2vec & word prediction &concatenation & location& Yes \\
     &network & Node2Vec &node prediction & \\
     &timestamp &  &NA& \\ 
      \bottomrule
    \end{tabular}
   \caption{Summary of User Embedding Methods}
   \label{tab:MethodOverview}
\end{table*}

\subsection{Text-based User Embedding}
The goal of text-based user embedding is to map a sequence of social media posts by the same user into a vector representation which captures the essential content and linguistic style expressed in the text. Here, we focus on methods that do not require any human annotations such as traditional unsupervised dimension reduction methods (e.g., {\em Latent Dirichlet Allocation} and {\em Single Value Decomposition}) or the more recent neural network-based prediction methods.

\subsubsection{Latent Dirichlet Allocation (LDA)} LDA is a generative graphical model that allows sets of observations to be explained by unobserved latent groups.  In natural language processing, LDA is frequently used to learn a set of topics (latent themes) from a large number of documents.  Several methods can be used to derive a user representation based on LDA results:(1)User-LDA which treats all the posts from each user as a single document and trains an LDA model to drive the topic distribution for this document. The per-document topic distribution is then used as the representation for this user. (2) Post-LDA which treats each post as a separate document and trains an LDA model to derive a topic distribution for each post. All the per-post topic distribution vectors by the same user are aggregated (e.g., by averaging them) to derive the representation of each user. According to~\cite{ding2017multi}, Post-LDA often learns better user representations than User-LDA in downstream applications. This may be due to the fact that social media posts are often short and thus each post may only have a single topic, which makes it  easier for LDA to uncover meaningful topics than from one big document containing all the user posts. 

\subsubsection{Matrix Representation}
Since we can also use a matrix to represent user-word and word-word co-occurrences, matrix optimization techniques are often used in learning user embeddings. If we use a sparse matrix to represent the relations between users and words where each row represents a user and each column represents a unique term in the dataset, we can use  matrix decomposition techniques such as Singular Value Decomposition (SVD) and Principle Component Analysis (PCA) to yield a set of more manageable and compact matrices that reveal hidden relations and structures in the data (e.g., correlation, orthogonality and sub-space relations). 



Recently, there is a surge of new text embedding methods that are designed to capture the semantics of words and documents. Except for \textbf {GloVe ( Global Vectors for Word Representation)},  which uses matrix optimization to learn a general representation of words, most text embedding methods employ neural network-based methods. Since neural network-based methods are supervised learning methods, to learn user embeddings, we often need an auxiliary training task for which a large number of training examples can be automatically constructed from raw social media data. We called these methods {\em self-supervised machine learning}.  


\subsubsection{Word Embedding}

Word2Vec is a popular neural network-based method designed to learn dense vector representations for individual words. The intuition behind the model is the Distributional Hypothesis, which states words that appear in the same context have similar meanings. There are two models for training a representation of word: Continuous Bag of Word (CBOW) and Skip Gram (SG) model. CBOW predicts a target word from one or more context words, while SG predicts one or more context words from a target word. Thus, predicting words in the neighborhood is the auxiliary task used to train word embeddings.  The models are frequently trained using either a hierarchical softmax function (HS) or negative sampling (NS) for efficiency. To learn user embeddings from social media posts, the word2vec model is first applied to learn a vector representation for each  word. Then a simple average of all the word vectors by the same user is used to represent a user~\cite{benton2016learning,ding2017multi}. 

GloVe is an unsupervised learning algorithm  designed to learn vector representations of words based on aggregated global word-word co-occurrence statistics from a text corpus. GloVe employs a global log bi-linear regression model that combines the advantages of global matrix factorization with that of local context window-based methods. GloVe has been used in  \cite{ding2017multi} to learn a dense vector for each word. To summarize all the words authored by a user,  we can use a vector aggregation function such as {\em average} to combine the vectors of all the words in a user's posts. 

\subsubsection{Document Embedding}
Doc2Vec is an extension of Word2Vec, which produces a dense low dimensional feature vector for a document.  There are two Doc2Vec models: Distributed Memory (DM) and Distributed Bag-of-Words (DBOW). Given a sequence of tokens in a document, DM can simultaneously learn a vector representation for each individual word token and a vector for the entire document. In DM, each sequence of words (e.g. a document) is mapped to a sequence vector (e.g., document vector) and each word is mapped to a unique word vector.  The document vector and one or more word vectors are aggregated to predict a target word in the context. DBOW learns a global vector to  predict tokens randomly sampled from a document. Unlike DM,  DBOW only learns a vector for the entire document. It does not use a local context window since the words for prediction are randomly sampled from the entire document. 

There are two typical methods for learning a user embedding from doc2vec results: (1) User-D2V which combines all the posts by
the same user in one document and trains a document vector to represent the user. (2) Post-D2V which treats each post as a document and train a doc2vec model to learn a vector for each post. To derive a user embedding,  all the post vectors from the same person can be aggregated using ``average”. 

\subsubsection{Recurrent Neural Networks (RNN).} 
The text embedding methods described above ignore the temporal  order of the words in a post and of the posts in a user account.  Since the order of text contains important information, to capture the sequential relations between words and posts, Recurrent Neural Network (RNN) models  such as  {\em Long Short-Term Memory (LSTM)} can be used~\cite{zhang2018user}. The input to an LSTM is a sequence of word embeddings and the output of an  LSTM is a sequence of hidden states, which are the input to downstream applications. Similar to word2vec, a language model-based auxiliary task is used to train a LSTM model on raw texts.  

Among all the text embedding methods we discussed, some employ prediction-based technologies (e.g., Word2Vec, Doc2Vec and LSTM), others use count-based methods (e.g., PCA, SVD, LDA and GloVe). There are some empirical evidence indicating that prediction-based methods may have some advantage over count-based methods in feature learning~\cite{baroni2014}. Among all the text embedding methods we discussed, only LDA generates embeddings that are somewhat interpretable. 


\subsection{Social Network-based User Embedding}
The objective of social network-based user embedding is to map very large social networks into low-dimensional embeddings that preserve local and global topological similarity. These methods focus primarily on learning a user representation that captures essential social structures and relations of a user.  The three most widely used network embedding methods are {\em DeepWalk}~\cite{perozzi2014deepwalk}, {\em Node2vec~\cite{grover2016node2vec}} and Matrix Factorization. 

\textbf{\em{DeepWalk}} learns latent
representations of vertices in a network from truncated random walks. It first generates short random walks. Each random walk  $S={v_{1}, v_{2},...,v_{l}}$  is treated as a sequence of words in a sentence.  DeepWalk then employs the SkipGram model (SG) in word2vec to learn the latent representation of a vertex. The learned embeddings can be used in many applications such as predicting user interests and anomaly detection~\cite{perozzi2014deepwalk}. 




\textbf{\em{Node2Vec}} is a modification of DeepWalk which employs a biased random walk to interpolate between Breadth-first Sampling (BFS) and Depth-first Sampling (DFS). With biased random walks, Node2vec can better preserve both the second-order and high-order proximity \cite{grover2016node2vec}. Given the set of neighboring  vertices generated by a biased random walk, Node2Vec learns the vertex representation using the SkipGram model (SG). The learned embedding has been used to characterize a Twitter user’s social network structure \cite{do2018twitter} and predict user interests~\cite{grover2016node2vec}.  

\textbf{\em Non-Negative Matrix Factorization (NMF)} is a matrix decomposition method with the additional constraint that all the entries in all the matrices have only positive values. The connections between network vertices are represented in an adjacency matrix.   Non-negative matrix factorization is used to obtain a low-dimensional embedding of the original matrix. NMF was  used in \cite{wang2017community} to learn  a network-based user embedding that preserves both the first- and second-order proximity. 


\subsection{Other Single-View User Embedding Methods}
In addition to texts and social networks, it is also possible to learn user embeddings from other types of social media data such as likes and images. For example, {\em User Like Embedding} was used in~\cite{kosinski2013private,ding2018predicting} for personality and delay discounting prediction. Many text-based user embedding methods are also applicable here. For example, SVD was used in ~\cite{kosinski2013private}; LDA, GloVe, Word2Vec, Doc2vec were used in ~\cite{ding2017multi,ding2018predicting}. In addition, {\em AutoEncoder (AE)} can be used in learning like embeddings. AE is a neural network-based feature learning method~\cite{autoencoder2006}. It learns an identity function so that the output is as close to the input as possible. Although an identity function seems a trivial function to learn, by placing additional constraints (e.g,, to make the number of neurons in the hidden layer much smaller than that of the input), we can still force the system to uncover latent structures in the data.  Finally, {\em Image-based User Embedding} can be obtained by extracting low-level latent image features from  pre-tained deep neural network models such as VGGNet~\cite{simonyan2014very}.

\section{Multi-View User Embedding}
To obtain a comprehensive and unified user representation based on all the social media data available, we need to combine user features from different views together. In addition to simply concatenating features extracted from different views, we can also apply machine learning algorithms to systematically fuse them. We categorize these fusion methods into two types: (a) general fusion methods (b) customized fusion methods. General fusion methods can be applied to diverse types of embedding vectors such as text and image embedding or text and like embedding  . In contrast, customized fusion methods are specifically designed to combine certain types of user data together. For example, ANRL is a method specifically designed to fuse user attributes and network topology together~\cite{zhang2018anrl}.

\subsection{General Fusion Methods}
First, we introduce two widely used general fusion methods. 

\textbf{\em {Canonical Correlation Analysis (CCA)}} CCA is a statistical method that explores the relationships between two multivariate sets of variables (vectors)~\cite{hardoon2004canonical}. Given two feature vectors,  CCA tries to find a linear transformation of each feature vector so that they are maximally correlated.  
CCA has been used in ~\cite{sharma2012generalized,ding2017multi} for multi-view fusion. 

\textbf{\em {Deep Canonical Correlation Analysis (DCCA)}}
DCCA is a non-linear extension of CCA, aiming to learn highly correlated deep architectures ~\cite{andrew2013deep}. The intuition is to find a maximally correlated representation of two feature vectors by passing them through multiple stacked layers of nonlinear transformation. Typically, there are three steps in training DCCA: (1) using a denoising autoencoder to pre-train each single view; (2) computing the gradient of the correlation of top-level representation; (3) tuning parameters using back propagation to optimize the total correlation.

The features learned from multiple views are often more informative than those from a single view. Comparing with single-view user embedding, multi-view embedding achieved significantly better performance in predicting demographics \cite{benton2016learning}, politic leaning ~\cite{benton2016learning} and substance use \cite{ding2017multi}.

\subsection{Customized Fusion Methods}
Several studies in our survey employ algorithms that are specifically designed to combine certain types of data. For example, \cite{zhang2017user} proposed a new algorithm called User Profile Preserving Social Network Embedding (UPPSNE), which combines user profiles and social network structures to learn a joint vector representation of a user. Similarly, Attributed Network Representation Learning (ANRL)~\cite{zhang2018user} employed a deep neural network to incorporate information from both network structure and node attributes. It  learns a single user representation that jointly optimizes AutoEncoder loss, SkipGram loss and Neighbour Prediction Loss.  \cite{liao2018attributed} proposed a  Social Network Embedding framework (SNE), which learns a combined representations for social media users by preserving both structural proximity and attribute proximity. \cite{ribeiro2018characterizing}  creates embeddings for each node with word embeddings learn from text using GloVe and the activity/network-centrality attributes associated with each user. So far, most of the customized fusion methods are designed to fuse network topology with additional node information (e.g., user profiles). 



\section{Embedding Fine Tuning Using Target Tasks}
In many cases, the learned user embeddings are simply used as the input to a target task. It is also possible that the learned user embeddings can be further refined to better support the target tasks with supervised learning.  For example,  in \cite{miura2017unifying}, the authors propose an attention-based neural network model to predict geo-location. It simultaneously learns text, network and metadata embeddings in supervised fine turning. In \cite{song2016volunteerism}, the authors collected multi-view user data from different platforms (e.g.,  Twitter, Facebook, LinkedIn  of the same user) and predicted volunteerism based on  user attributes and network features. Then,  they combine these two sets of features in supervised fine tuning to enhance the final prediction. \cite{farnadi2018user} learned a hybrid user profile which is a shared user representation learned from three data sources. They were combined at decision level to predict multiple user attributes (e.g., age, gender and personality).

\section{Discussion}
Large-scale social media-based user trait and behavior analysis is an emerging multidisciplinary field with the potential to transform human trait and behavior analysis from controlled small scale experiments to large scale studies of natural human behavior in an open environment. Although raw social media data are relatively easy to obtain, it is  expensive to acquire the ground truth data at a large scale. The proposed unsupervised/self-supervised user embedding methods can alleviate the ``small data" problem by transferring knowledge learned from raw social media data to a new target task. This step is very important for many human trait and behavior analysis applications. According to ~\cite{benton2016learning,preoctiuc2015analysis}, machine learning models that incorporate unsupervised/self-supervised user embedding significantly outperform the baselines that do not employ use embedding. Based on the survey, we have also identified a few major issues in the current social media analysis research. 
\subsection{Interpretability}
Although systems employing user embeddings significantly outperform baselines in terms of prediction accuracy, these systems also suffer from one significant drawback: low interpretability. Since  user embedding features are latent features automatically uncovered by the system, it is often difficult for humans to understand the meaning of these features. This  may significantly impact our ability to gain insight into these behavioral models. So far, there have not been much work focusing on learning user representations that are both effective and interpretable.     

\subsection{Ethical Issues} Due to the privacy concerns in accessing user data on social media and the sensitive nature of the inferred user characteristics, if not careful, there could be significant privacy consequences and ethical implications. So far,  most of the studies in our survey focused primarily on  optimizing system performance. There have not been sufficient discussion on ethical concerns when conducting research in this field. 

\section{Future Directions}
Each of the main issues we identified above also presents a good opportunity for future work.  

\subsection{Interpretable User Representation Learning} we need more research on learning high-performance user representations that are also interpretable. Some preliminary work has conducted in this area.  In~\cite{ding2018interpreting}, a knowledge distillation framework was proposed to train behavior models that are both highly accurate and  interpretable. Developing causal models for both inference and interpretation is another potential new direction. 

\subsection{Ethical Research on Data-driven Behavior Analysis } Ethical issues are complex, multifaceted and resist simple solutions. In addition to privacy concerns  in data collection, researchers working on social media-based human trait and behavior analysis also face  other ethical challenges including informed consent, traceability and working with children and young people. There is an urgent need for the research community to decide an ethical framework to  guide researchers to navigate obstacles, gain trust and still allow them to capture salient behavioral and social phenomena.  Recently there is a surge of interests and research on fair data-driven decision making. As a researcher, we also need to be aware of the potential impact of social media analytics on the well-being of individuals and our society.       

We have also identified a few possible research directions to improve the state of the art user embedding techniques.     

\subsection{Temporal User Embedding} since most social media data are  collected over a long period of time and associated with time stamps, it is an ideal data source for longitudinal data analysis. Also, for many medical and public health applications, analyzing  behavioral changes over time is critical to understanding one's decision making process. Although Recurrent Neural Networks such as LSTM can capture some sequential patterns, they totally ignore the time stamp associated with each event. More work on learning user embedding from time is needed. 

\subsection{User Embedding with Multi-task Learning} Since individual traits and behavior are highly correlated, building a prediction model that simultaneous infer multiple correlated traits and behavior should yield better performance than predicting each trait/behavior separately. Most existing studies only predict one user attribute/behavior at a time. More research is  needed to jointly train and predict multiple user attributes together for better performance. 

\subsection{Cross-platform Fusion} It is also common for a user to have multiple accounts on different social media platforms. Recently, new technologies have been developed to link different social media accounts of the same user together~\cite{abel2013cross}. With this linked data, it is possible to perform novel cross-platform user trait and behavior analysis such as (1) domain bias analysis that focuses on studying the impact of domain or social media platform on user trait and behavior analysis, (2) domain adaptation that addresses how to adjust prediction models trained on one platform (e.g., Twitter) to predict the traits and behavior on another platform (e.g., Facebook). So far, there is some initial work on domain bias analysis and correction~\cite{kilicc2016analyzing}. More research is needed in order to develop more robust tools for human trait and behavior analysis.



\clearpage
{\footnotesize
\bibliographystyle{named}
\bibliography{ijcai19}
}

\end{document}